\documentclass{aastex631}

\shorttitle{No Pulsar Companion Around the Nearest Low Mass White Dwarf}
\shortauthors{Athanasiadis et al.}
\graphicspath{{./}{figures/}}

\begin{document}
\title{No Pulsar Companion Around the Nearest Low Mass White Dwarf}

\author{Tilemachos M. Athanasiadis}
\affiliation{Max-Planck-Institut f\"{u}r Radioastronomie\\
Auf dem H\"{u}gel 69, 53121 \\
Bonn, Germany}

\author{Nataliya K. Porayko}
\affiliation{Max-Planck-Institut f\"{u}r Radioastronomie\\
Auf dem H\"{u}gel 69, 53121 \\
Bonn, Germany}

\author{John Antoniadis}
\affiliation{Institute of Astrophysics, FORTH, Dept. of Physics, University of Crete\\
Voutes, University Campus, GR-71003 \\
Heraklion, Greece}
\affiliation{Max-Planck-Institut f\"{u}r Radioastronomie\\
Auf dem H\"{u}gel 69, 53121 \\
Bonn, Germany}

\author{David Champion}
\affiliation{Max-Planck-Institut f\"{u}r Radioastronomie\\
Auf dem H\"{u}gel 69, 53121 \\
Bonn, Germany}

\author{Olaf Wucknitz}
\affiliation{Max-Planck-Institut f\"{u}r Radioastronomie\\
Auf dem H\"{u}gel 69, 53121 \\
Bonn, Germany}

\author{Benedetta Ciardi}
\affiliation{Max-Planck-Institut f\"{u}r Astrophysik\\
Karl-Schwarzschild-Strasse 1, 85748 \\
Garching, Germany}

\author{Matthias Hoeft}
\affiliation{Th\"uringer Landessternwarte, \\
Sternwarte 5, 07778 Tautenburg, Germany}

\author{Michael Kramer}
\affiliation{Max-Planck-Institut f\"{u}r Radioastronomie\\
Auf dem H\"{u}gel 69, 53121 \\
Bonn, Germany}

\begin{abstract}

2MASS J050051.85$-$093054.9 is the closest known low-mass helium-core white dwarf in a binary system. We used three high-band international LOFAR stations to perform a targeted search for a pulsar companion, reaching sensitivities of $\sim$\,3\,mJy for a 10-ms pulsar at a DM = 1 pc\,cm$^{-3}$. No pulsed signal was detected, confidently excluding the presence of a detectable radio pulsar in the system. 

\end{abstract}

\keywords{stars: white dwarfs --- pulsars: general --- survey }

\section{Background} \label{sec:mot}

Recently, \citet{2020kawka} reported that the binary system 2MASS J050051.85$-$093054.9 (hereafter J0500-0930) hosts the extremely low-mass dwarf (LMWD) which is closest to the Sun, at a distance of $\sim$ 71 pc.
The system has an orbital period of $\sim$9.5\,h and the mass of the white dwarf (WD) is $m_{\rm WD}=0.17\pm0.01$\,M$_{\odot}$. 
The system's mass function is $ f \equiv {P_{\rm orb}\,{K^3}}/{2\pi G}=0.12$\,M$_{\odot}$, 
where $P_{\rm orb}$ is the orbital period and $K$ the semi-amplitude of the WD orbital radial velocity. 
Based on the latter, and assuming an inclination distribution that is uniform in $\cos i$, the probability for the system to host a neutron star (NS) with a mass $\geq 1.3$\,M$_{\odot}$ is 6\%  \citep[see][]{2021Athanasiadis}. 
If the binary companion is indeed a NS, then the system formed via long-term mass transfer in a low-mass X-ray binary  \citep{1991Bhattacharya}. The NS would be spun-up due to gain of angular momentum and thus could potentially be observable as a rapidly-spinning millisecond pulsar (MSP). 
Only a few MSP/LMWD systems are known so far, and thus a possible discovery at that short distance would be extremely valuable \citep{Antoniadis:2014eia, Wex:2020ald}.

Assuming a standard luminosity function \citep[e.g][]{Faucher-Giguere:2005dxp}, the expected flux density for an MSP at this distance is $\gtrsim 1$\,Jy (see Figure~\ref{fig:det_plot}).  The expected dispersion measure (DM) based on the \citet{NE2001} model for the Galactic free electron distribution is $\le 1$ pc\,cm$^{-3}$.

\section{Observations, analysis and results} \label{sec:obs}
\label{observations}
The estimated low DM of the system is of particular concern with respect to searching for pulsar signals. 
For radio receivers with a typical $\sim$200\,MHz bandwidth operating at L-band, the time delay due to the dispersion between the lower and upper parts of the band would  amount to only few percent of the MSP period. 
In this case, a pulsar detection becomes challenging as the potential signal is potentially difficult
to distinguish from RFI, usually showing up as a spurious signal at DM $\sim$\,0 pc\,cm$^{-3}$. 
Motivated by this issue, we have strategically shifted the observational setup for the search campaign to the lowest radio frequencies available, where the dispersion sweep of the signal can be maximised.

For this purpose, we used three of the German stations in the the Low-Frequency Array \citep[LOFAR; ][]{2013vanHaarlem} to conduct a sensitive search for periodic radio pulses from this binary. More specifically, 
the system was observed on April 28, 2020 from 14:30 up to 15:30 UTC  
using the DE601 (Effelsberg), DE602 (Unterweilenbach near Munich) and DE603 (Tautenburg) stations.  
The observations were performed with the high band antennas (HBA) in the second Nyquist zone, from 102.25 to 197.56 MHz (\texttt{RCU mode 5}, split in 4 lanes). The analysis treated the resulting datasets independently. 
A coherent combination of the signals from all three stations would require delay and phase calibration. 
This was not possible due to the lack of additional calibrator scans and because no signal from the target was detected. 
Incoherent or semi-coherent addition would not offer the same sensitivity benefit therefore we decided to keep the data separate for consistency checks.
Considering the effective collective area of the GLOW stations, their gain dependency on azimuth and elevation \citep[see][]{2013Haarlem,2016Kondratiev}, the sensitivity of each individual observation per station, for a 10-ms pulsar at $\textrm{DM}=1$\,pc\,cm$^{-3}$, was  $\sim$ 5\,mJy. 

For the data analysis we used a hybrid \texttt{SIGPROC}- \citep{sigproc2011} and \texttt{PRESTO}- based pipeline \citep{Ransom2001} on MPIfR's \textsc{hercules} cluster at the Max-Planck Computing and Data Facility 
(MPCDF) in Garching.
Raw baseband data from each station were converted to filterbank format with 1952 channels, and a time resolution of 81.92 $\mu$s.
The filterbank spectra were then de-dispersed at 100 trial DM values ranging from 0 to 5 pc\,cm$^{-3}$. 

For the mitigation of the radio frequency interference (RFI) from the de-dispersed timeseries, the \texttt{RFIFIND} routine of \texttt{PRESTO} was used. 
The clean de-dispersed time series were then searched for periodic signals (from 1\,ms up to 10 seconds) using fast Fourier transforms (FFT) and incoherent summing of 16 harmonics.
A method known as acceleration search was also used in order to account for the possible smearing of signals
in the Fourier 
bins of the power spectrum due to the orbital motion. 
Following this, we re-sampled the time series assuming constant acceleration within the range of $\pm$50\,${\rm m}/{\rm s}^2$ which is more than sufficient, given the orbital characteristics of the system.

\begin{figure}
    \centering
    \includegraphics[width=0.9\columnwidth]{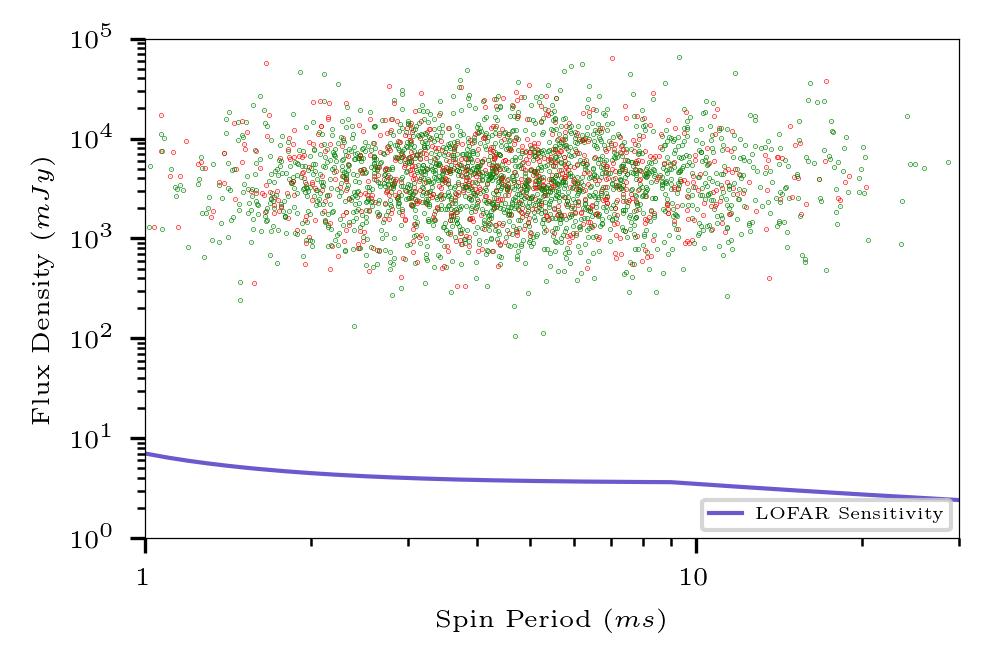}
    \caption{Estimates for the expected pulsar flux density at the frequency of our observations, assuming the luminosity function of \cite{Faucher-Giguere:2005dxp} and a spectral index of $-1.8$ \citep[green points; see][for details]{2021Athanasiadis}. Red points correspond to pulsars that are beamed away from Earth, assuming a beaming fraction of 70\% \citep{Kramer:1998abc}. The blue line corresponds to a conservative sensitivity limit of our observations at DM = 1\,${\rm pc}\,{\rm cm}^{-3}$.}
    \label{fig:det_plot}
\end{figure}

Significant candidates (with signal-to-noise ratio $\geq$7) were examined by eye, but no robust pulsar-like signal was identified. 
Given the sensitivity of these observations, this confidently excludes the presence of a radio pulsar in the system (with beam directed towards Earth, see Figure~\ref{fig:det_plot}).

\begin{acknowledgments}
This work is based on data obtained with the German stations of the International LOFAR Telescope (ILT), constructed by ASTRON \citep{2016Kondratiev}, during station-owners time. In this work we made use of data from the Effelsberg (DE601) LOFAR station funded by the Max-Planck-Gesellschaft; the Unterweilenbach (DE602) LOFAR station funded by the Max-Planck-Institut f\"{u}r Astrophysik, Garching; and the Tautenburg (DE603) LOFAR station funded by the State of Thuringia, supported by the European Union (EFRE) and the Federal Ministry of Education and Research (BMBF) Verbundforschung project D-LOFAR I (grant 05A08ST1).
The observations of the German LOFAR stations were carried out in the stand-alone GLOW mode (German LOng-Wavelength array), which is technically operated and supported by the Max-Planck-Institut f\"{u}r Radioastronomie, the Forschungszentrum J\"{u}lich and Bielefeld University. We acknowledge support and operation of the GLOW network, computing and storage facilities by the FZ-J\"{u}lich, the MPIfR and Bielefeld University and financial support from BMBF D-LOFAR III (grant 05A14PBA), D-LOFAR IV (grant 05A17PBA) and D-LOFAR2.0 (05A20STA), and by the states of Nordrhein-Westfalia and Hamburg. \\
We would like to thank Tobia Carozzi and Jompoj Wongphecauxson for discussions. JA acknowledges support by the Stavros Niarchos Foundation (SNF) and the Hellenic Foundation for Research and Innovation (H.F.R.I.) under the 2nd Call of ``Science and Society'' Action Always strive for excellence -- ``Theodoros Papazoglou'' (Project Number: 01431) 
\end{acknowledgments}

\vspace{5mm}
\facilities{LOFAR, MPIfR's \textsc{hercules} cluster}

\software{Astropy \citep{2013A&A...558A..33A,2018AJ....156..123A}, 
          NumPy \citep{harris2020array}, 
          Matplotlib \citep{Hunter:2007}
          }

\bibliography{sample631}{}
\bibliographystyle{aasjournal}

\end{document}